\documentclass[prl,twocolumn,showpacs,preprintnumbers,amsmath,amssymb]{revtex4-1}

\usepackage{graphicx}
\usepackage{dcolumn}
\usepackage{bm}
\usepackage{amssymb}
\usepackage{amsmath} 
\usepackage{wasysym}
\usepackage{sidecap}
\usepackage{natbib} 

\def\kpara{\textit{{k}$_\parallel$}}

\def\invA{\AA$^{-1}$}
\def\Kbar{$\overline{\rm K}$}

\def\GbarMbar{$\overline{\Gamma \rm M}$}
\def\GbarKbar{$\overline{\Gamma \rm K}$}
\def\GbarMbarBlueP{$\Gamma \rm M$}
\def\GbarKbarBlueP{$\Gamma \rm K$}
\def\KbarMbarBlueP{$\rm K \rm M$}

\begin{document}

\title{Band renormalization of blue phosphorus on Au(111)}

\author{E. Golias$^{1,2}$}
\email{evangelos.golias@gmail.com}
\author{M. Krivenkov$^1$}
\author{A. Varykhalov$^1$}
\author{J. S\'anchez-Barriga$^1$}
\author{O. Rader$^1$}
\affiliation{$^1$Helmholtz-Zentrum Berlin f\"ur Materialien und Energie, Elektronenspeicherring BESSY II, Albert-Einstein Stra{\ss}e 15, 12489 Berlin, Germany}
\affiliation{$^2$Institut f\"ur Experimentalphysik, Freie Universit\"at Berlin, Arnimallee 14, 14195 Berlin, Germany}

\begin{abstract}
Most recently, theoretical calculations predicted the stability of a novel two-dimensional phosphorus honeycomb lattice named blue phosphorus. Here, we report on the growth of blue phosphorus on Au(111) and unravel its structural details using diffraction, microscopy and theoretical calculations. Most importantly, by utilizing angle-resolved photoemission spectroscopy we identify its momentum-resolved electronic structure. We find that Au(111) breaks the sublattice symmetry of blue phosphorus leading to an orbital-dependent band renormalization upon the formation of a (4x4) superstructure. Notably, the semiconducting two-dimensional phosphorus realizes its valence band maximum at 0.9 eV binding energy, however, shifted in momentum space due to the substrate-induced band renormalization.
\end{abstract}

\pacs{73.20.At, 73.20.-r, 79.60.-i, 79.60.Bm}

\maketitle

The advent of graphene\cite{Novoselov:2004it} led to the discovery of novel two-dimensional material platforms opening the way for the investigation of intriguing physical phenomena that promise numerous technological applications. The ever-growing list of low-dimensional materials, leaving aside the vast family of transition metal dichalcogenides\cite{Wang:2012fa, Jariwala:2014ih, Manzeli:2017ib}, includes the two-dimensional mono-elemental lattices of silicene\cite{Vogt:2012cf}, germanene\cite{Acun:2015ia}, stanene\cite{Zhu:2015hc} and borophene\cite{Mannix1513, Feng:2017hn}. Recently, phosphorus joined the two-dimensional world with the rediscovery of an allotrope of phosphorus, namely black phosphorus\cite{Churchill:2014fa,Ling:2015ba,Golias:2016cg} , which has already attracted a lot of attention mainly because of its excellent performance in optoelectronic devices\cite{Li:2014gf, Xia:2014cq,Liu:2014kc}. Most recently, theoretical calculations suggested the stability of another phosphorus allotrope, named blue phosphorus (BlueP)\cite{Zhu:2014kf}, which comprises a buckled honeycomb lattice. 

Since BlueP does not exist in nature we have to engineer it on a suitable substrate. Noble metals have played an important role in the growth of materials with similar two-dimensional structure such as silicene and germanene. Several two-dimensional phases of Si and Ge have been realized on different metal surfaces\cite{Vogt:2012cf, Acun:2015ia, Meng:2013kh, Li:2014ed}, however, the strong interaction with their supporting substrates\cite{Tsoutsou:2013bi, Golias:2013gy} hindered the emergence of the expected graphene-like electronic properties based on theoretical predictions\cite{Cahangirov:2009fh}. Likewise, the formation of BlueP on Au(111) was recently reported\cite{Zhang:2016ex, Xu:2017fg, Gu:2017ko}, nevertheless, its electronic structure has yet to be observed. A direct comparison between the experimentally observed electronic band structure with theoretical predictions is the key to discern BlueP from a substrate-induced two-dimensional phosphorus phase.

In the present Letter, we report on the momentum-resolved electronic structure of BlueP on Au(111). As a first step, we pinpoint the structural details of the observed BlueP phase on Au(111) by combining low-energy electron diffraction (LEED), scanning tunneling microscopy (STM) and density functional theory (DFT) calculations. We unambiguously identify a (5x5) superstructure on Au(111), which is commensurate with a (4x4) BlueP lattice, and refine its structural details using DFT. Later on, we utilize angle-resolved photoemission spectroscopy (ARPES) to identify the electronic signature of (4x4)-BlueP on Au(111), link its experimental band structure with a DFT-refined model and unravel its relation with freestanding BlueP.

The photoemission measurements were conducted at the ARPES $1^2$ end-station of the U112-PGM2a beamline of BESSY II using a Scienta R8000 hemispherical electron analyzer at ambient temperature. The STM measurements were performed with an Omicron VT STM, operated at room temperature. During ARPES and STM measurements the base pressure of both chambers was better than $2\times 10^{-10}$ mbar. A clean and sharp Au(111) surface was prepared by repeated cycles of Ar$^+$ sputtering (ion energy 1 keV, pressure $1\times 10^{-5}$ mbar) followed by annealing at 620 $^{\circ}$C for 5 mins and subsequently at 420 $^{\circ}$C for 20 mins. For the BlueP growth we used a black phosphorus crystal as precursor, heated to 300 $^{\circ}$C and placed facing the clean Au(111) surface at approximately 4 cm distance. We prepared and characterized two different samples, one with the Au(111) substrate held at 230 $^{\circ}$C while depositing P and another where P was deposited at room temperature and then the sample was annealed at 250 $^{\circ}$C for 15 mins.

DFT calculations were performed using projector-augmented wave potentials \cite{Blochl:1994paw} and the exchange-correlation functional of Perdew, Burke and Ernzerhof \cite{pbe:1996pp, *pbe:1996er} as implemented in the VASP package \cite{kresse:1999vas}. To simulate BlueP we employed a plane-wave basis with kinetic energies up to 400 eV for the expansion of the electronic wavefunctions while a $\Gamma$-centered reciprocal grid of 15$\times$15 k-points was used for the basal plane sampling. The equilibrium configuration was found by scanning the lattice constant of the hexagonal freestanding BlueP lattice while relaxing its atomic positions until the energy change between consecutive minimization steps and the maximum force on atoms were $1 \times 10^{-6}$ eV and 0.001 eV/\AA, respectively.  A slab configuration incorporating four gold layers, the phosphorus overlayer and a vacuum region of 18 \AA\ was utilized for the simulation of the (4x4)-BlueP on Au(111), the reciprocal space was sampled by a 9$\times$9 k-point grid. The three bottom layers were fixed at their experimental bulk positions while the top Au layer along with the phosphorus adlayer, comprising 32 atoms, were allowed to relax using a conjugate-gradient algorithm until the forces on non-fixed atoms were lower than 0.01 eV/\AA\ .

\begin{figure}
\includegraphics [width=0.48\textwidth]{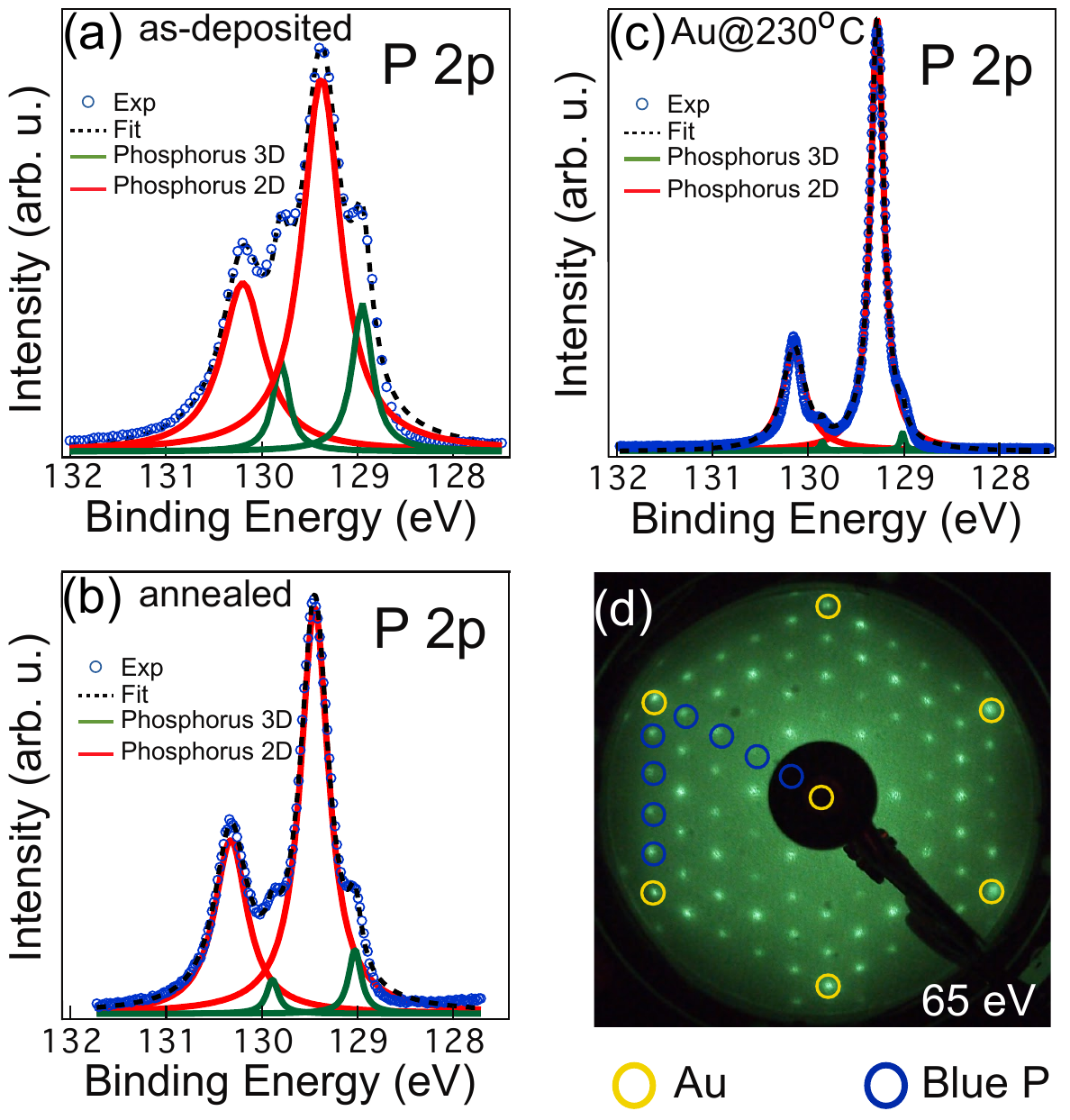}
\caption{XPS spectra of the phosphorus 2p core-level of the (a) as-deposited phosphorus on Au(111),  (b) after annealing at 250 $^{\circ}$C for 15 minutes and (c) after depositing P on Au(111) heated at 230 $^{\circ}$C, respectively. (d) LEED (5x5) diffraction pattern --with respect to Au(111)-- of (4x4)-BlueP on Au(111) measured with electrons of $E_{kin} = $ 65 eV. Gold and blue circles highlight diffraction spots corresponding to (1x1)-Au(111) and (4x4)-BlueP, respectively.}
\label{fig1}
\end{figure}

The formation of BlueP on Au(111) starts with the sublimation of the precursor in the form of P$_2$ molecules\cite{fortin:2016sub, Liu:2016sub}, which condense on the Au(111) surface. A post-annealing step or setting the substrate at elevated temperature provides the required energy to the P atoms for the formation of a self-limiting\cite{Xu:2017fg} honeycomb monolayer which is energetically favored on a substrate with hexagonal symmetry\cite{Zhu:2014kf}. Both methods lead to a monolayer BlueP with high coverage on Au(111), nevertheless, growing BlueP while keeping the substate at elevated temperature results to a BlueP layer with larger domain size\cite{Zhang:2016ex}. In Fig. \ref{fig1}(a), (b) and (c) we present the 2p core-level doublet of as-deposited P held at room temperature, after a subsequent annealing step and of P deposited on Au(111) substrate held at 230 $^{\circ}$C, respectively. The P 2p core-level in Fig. \ref{fig1}(a) comprises two components which correspond to a two- (interface) and three-dimensional (bulk) contribution. Phosphorus atoms proximate to the Au(111) surface give rise to the higher binding energy doublet similar to what has been observed during Ge growth on Au(111)\cite{Davila:2016gb, Schroter:2017kj}. After annealing (see Fig. \ref{fig1} (b)) the bulk-like component of the P 2p core-level drastically reduces, while the interface component dominates, indicating the formation of a two-dimensional structure on Au(111). When the Au substrate is set to 230 $^{\circ}$C during P deposition the three-dimensional component virtually vanishes, an indication of a more well defined two-dimensional P layer, which is further supported by the smaller linewidth of the two-dimensional doublet due to the reduction of defects\cite{Genesan:2016xps}. Structurally, the recorded LEED diffraction patterns shown in Fig. \ref{fig1} (d) reveal a well-defined (5x5) superstructure with respect to the Au(111) surface, which approximately matches four times the periodicity of BlueP if one considers its theoretically predicted lattice constant in the current work. The observed evolution of the P core levels along with the structural information from diffraction corroborates the proposed growth mechanism of monolayer BlueP on Au(111) surface.

 \begin{figure}
 \includegraphics [width=0.49\textwidth]{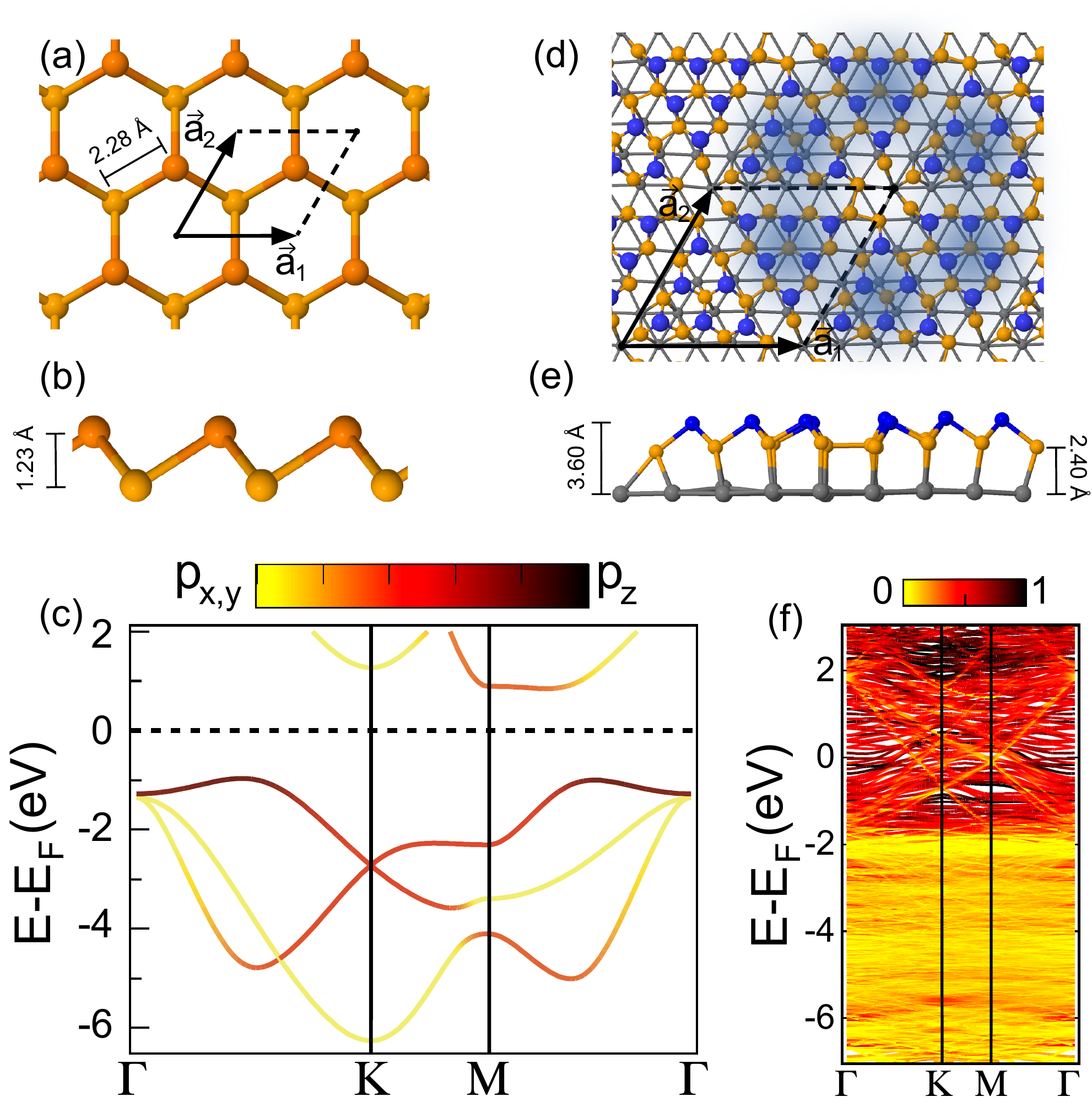}
 \caption{Ball and stick model as viewed from (a) top and (b) side of the fully relaxed freestanding BlueP. Dark and light orange spheres correspond to phosphorus atoms lying in two different planes in the freestanding BlueP, respectively. (c) Band structure of freestanding BlueP. The color scale highlights the in- and out-of-plane orbital contribution to the electronic bands. (d) Top and (e) side view of the relaxed (4x4)-BlueP on top of (5x5)-Au(111). Blue, orange and grey spheres correspond to high-, low-buckling phosphorus and gold atoms, respectively. (f) Electronic structure of the configuration presented in (d), (e), where the color scale represents the relative contribution to the band structure of the electronic states projected on the phosphorus atoms.}
 \label{fig2}
 \end{figure}
The equilibrium configuration of freestanding BlueP is depicted in Fig. \ref{fig2} (a), (b). BlueP forms a buckled honeycomb structure with a lattice constant of 3.33 \AA, distance between the two inequivalent atomic planes of 1.23 \AA\ and a P-P bond of 2.28 \AA. In Fig. 2 (c) we present the band structure of freestanding BlueP along with the relative contributions of in- and out-of-plane p-orbitals that dominate the electronic bands of the material in the proximity of the Fermi level. The top and bottom of the valence band (VB) and conduction band (CB), respectively, have a strong p$_z$ character and give rise to an indirect band gap of about 1.81 eV, given the well-known tendency of DFT calculations to underestimate the band gaps of insulators and semiconductors\cite{Perdew:2017gap}. 
More specifically, the top of the VB is located away from the $\Gamma$ point of the surface Brillouin zone (BZ) along the $\Gamma \rm K$ high-symmetry direction approximately 0.9 eV below the Fermi level, while the bottom of the CB is located along the $\Gamma \rm M$ direction, thereby, BlueP is a indirect gap semiconductor.
In order to find a more realistic model for BlueP on Au(111) and based on the diffraction information of Fig. \ref{fig1} (d) we let a (4x4) freestanding BlueP structure to relax on top of a (5x5) Au(111) surface. The relaxed unit cell of the (4x4)-BlueP (highlighted with black lines in Fig. \ref{fig2}(d), (e)) encompasses two different groups of phosphorus atoms: twenty are closer to the substrate located mainly above on-top and bridge sites of the Au(111) surface while twelve atoms occupy buckled positions that predominantly project onto the vicinity of the hollow sites of the Au(111) surface. These two atomic groups lie in different levels with an average distance above the Au(111) surface of 2.40$\pm$0.17 \AA\ and 3.60$\pm$0.14 \AA, respectively, resulting in a distance between low- and high-lying phosphorus atoms that virtually matches the buckling height of the ideal freestanding BlueP. Overall the relaxed (4x4)-BlueP configuration (see Fig. \ref{fig2} (d)) resembles a flower-like shape, which differs from the highly-symmetric atomic arrangement of BlueP, with its unit cell comprising two triangular shaped regions where the low- and high-buckled atoms reduce the average P-P bond by 7.8\% (2.26$\pm$0.22 \AA) compared to the freestanding configuration of Fig. \ref{fig2}(a). Notably, within the triangular segments of the (4x4) unit cell the atomic arrangement is similar to the ideal BlueP and only a trench-like region separating these two segments differentiates the (4x4)-BlueP from freestanding BlueP. On the other hand, the tensile strain on the superstructure amounts to 8.1\% if one considers the freestanding lattice constant of BlueP on Au(111). The band structure of Fig. \ref{fig2} (d) projected onto the P atoms, and presented in Fig \ref{fig2} (f), indicates that one expects to observe P-related bands above the region which is dominated by the gold d-bands, i.e. between 2 eV of binding energy (BE) and the Fermi level.
\begin{figure}
\includegraphics [width=0.49\textwidth]{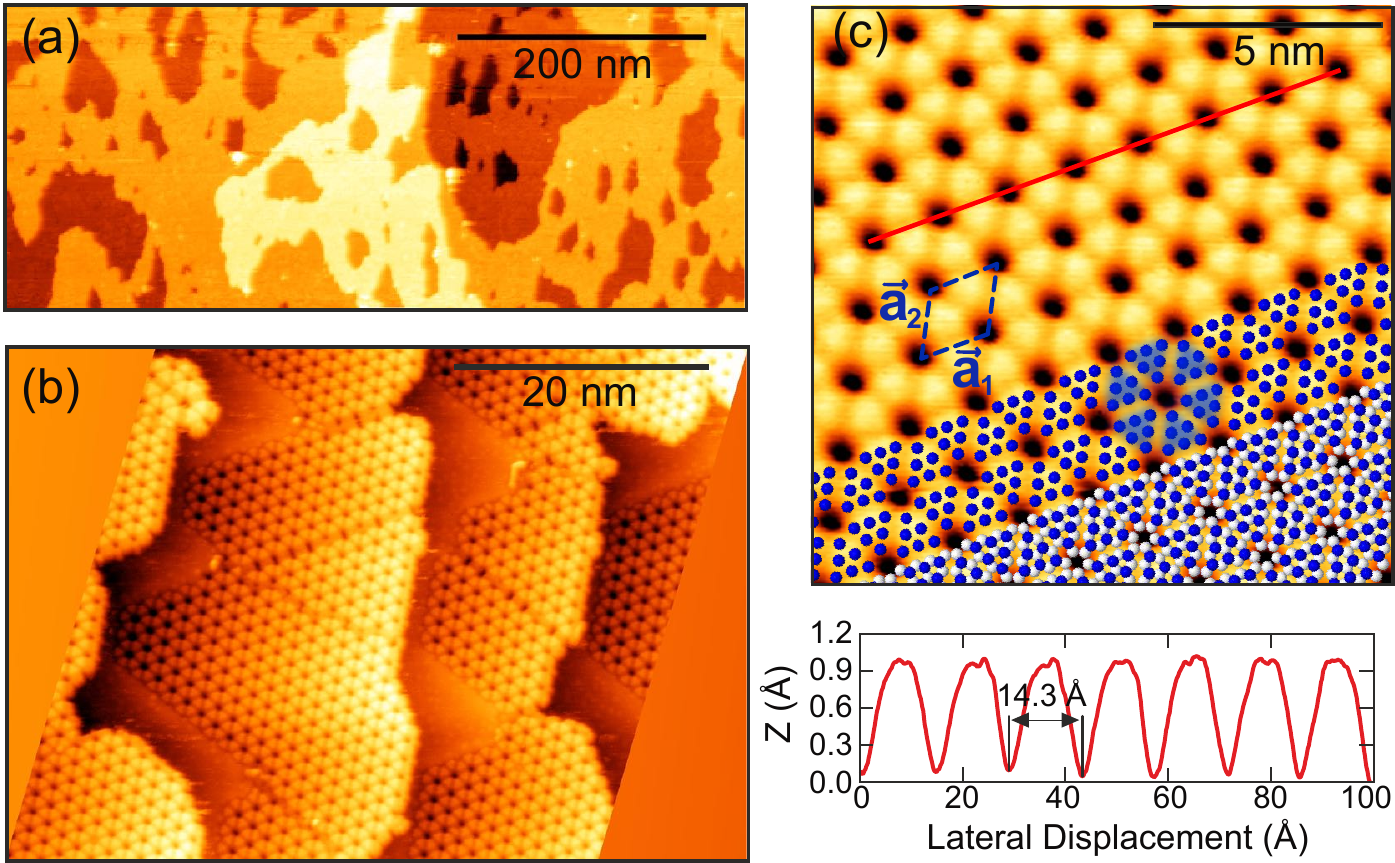}
 \caption{STM characterization of (4x4)-BlueP on Au(111). Corresponding bias voltages, tunneling currents  and scanning areas are given in brackets. (a) Large scale image, one can see a rough edge morphology of P overlayer and its high surface coverage [1V, 0.1 nA, 593$\times$244 nm$^2$]. (b) Au(111) terraces with single layer P that tends to be attached to the Au terrace edges [1V, 0.6 nA, 62$\times$41 nm$^2$]. (c) Close-up image with the DFT optimized atomic structure overlaid [1V, 1.52 nA, 13$\times$13 nm$^2$]. Blue and white balls represent high- and low-buckled P atoms, respectively. Low-buckled phosphorus layer was removed for clarity in the upper part of the overlay. Triangles in the STM images correspond to groups of six high-buckled phosphorus atoms. Six triangles form a hexagon (highlighted with blue) around a deep in the center. Unit cell of (4x4)-BlueP is marked with blue dashed line. The height profile of the structure taken along the red line can be seen below.}
 \label{fig3}
 \end{figure}
 
STM topographical scans recorded after the annealing of the sample are presented in Fig. \ref{fig3}. Figure \ref{fig3} (a) depicts the P-covered Au(111) surface where brighter colors correspond to topographically higher regions. Based on the STM measurements we can conclude that nearly the whole Au(111) surface is covered with BlueP patches. A zoom-in on the covered regions, presented in Fig. \ref{fig3} (b), reveals a planar honeycomb-like structure occupying the Au(111) terraces. Figure \ref{fig3} (c) reveals finer details of the (4x4)-BlueP structure on Au(111), where the flower-like shape predicted by the DFT calculations, presented in Fig. \ref{fig2} (d) and (e), becomes apparent when the high-buckled P atoms form the observed triangular segments. The overlaid DFT-optimized structure in the bottom of the STM image shows a remarkable agreement between STM and DFT calculations. We have to note that we employed  molecular dynamics simulations to search for a two-dimensional configuration with lower energy compared to the one of Fig. \ref{fig2} (d), (e), however, we did not find a structure more energetically stable which complies with our STM observations. From the height profile along the red line depicted at Fig. \ref{fig3} (c) we measured the distance between two dark round regions, i.e one lattice constant of the (4x4)-BlueP superstructure, to be 14.3 \AA\ very close to five times the lattice constant of Au(111) (14.45 \AA), which is also consistent with the diffraction pattern of Figs. \ref{fig1} (d). Finally, we observe that in contrast to graphene, which forms a quasi-freestanding layer and extends over the step edges on most supporting substrates, P interacts strongly with Au(111) and even bonds to the step edges. In our microscopy measurements we were even able to distinguish small patches of BlueP incorporated into the top-most Au layer (see Supplementary information\cite{Supplementary}).

\begin{figure*}
\includegraphics [width=0.90\textwidth]{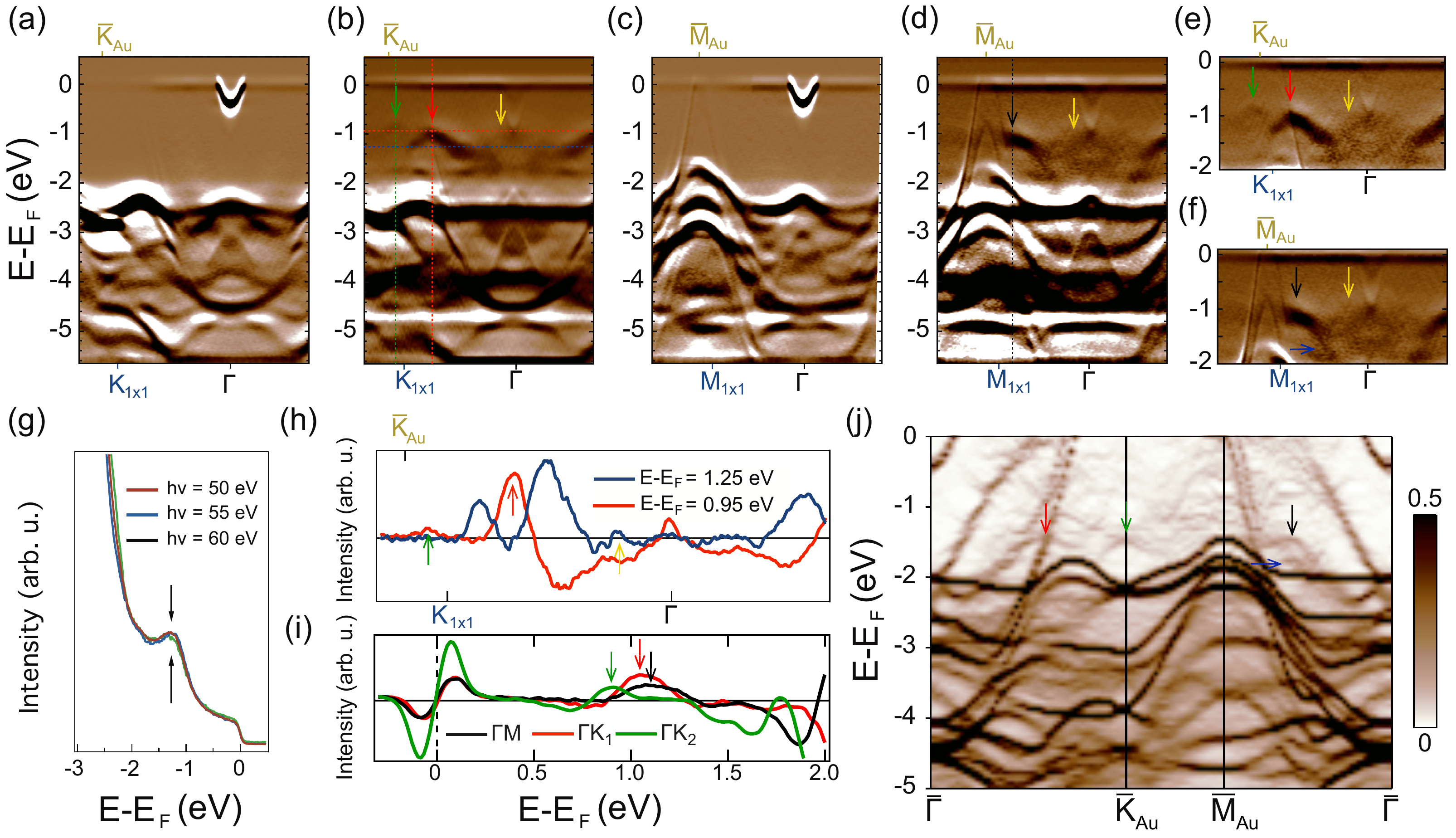}
\caption{Second derivative plots of the ARPES spectra for (a) clean Au(111) and (b) (4x4)-BlueP on Au(111) along the $\Gamma \rm K$, respectively, using horizontally polarized light with h$\nu = 60$ eV. Second derivative plots of the ARPES spectra for (c) clean Au(111) and (d) BlueP on Au(111) along the $\Gamma \rm M$, respectively, using horizontally polarized light with of h$\nu = 55$ eV. (e), (f) Close-up in the energy region from Fermi level down to 2 eV BE for the spectra of (b) and (d), respectively. (g) Energy distribution curves at \kpara\ = 0.8 \invA\ for the BlueP on Au(111) along the \GbarKbarBlueP\ high-symmetry direction for three different excitation energies. (h) Momentum distribution curves extracted from (b) at 0.95 and 1.25 eV BE, marked by red and blue dashed lines in (b), respectively. (i) Red and green lines correspond to energy distribution curves at \kpara\ = 0.95 \invA\ and \kpara\ = 1.31 \invA\ along the \GbarKbarBlueP\ direction, respectively, extracted from (b). The black line corresponds to the energy distribution curve along the \GbarMbarBlueP\ high-symmetry direction of (d) at \kpara\ = 0.92 \invA. The momentum positions of the extracted energy distribution curves are marked with the corresponding dashed line color in (b) and (d).  (j) Effective band structure of (4x4)-BlueP on Au(111) as presented in Fig. \ref{fig2}(f) unfolded to the primitive cell of Au(111) surface. In the subfigures, gold and blue letters mark the high-symmetry points of Au(111) and freestanding BlueP, respectively. Distances of high-symmetry segments in the reciprocal space for BlueP are \GbarKbarBlueP=1.26 \invA, \GbarMbarBlueP=1.09 \invA, and for Au(111) \GbarKbar=1.44 \invA, and \GbarMbar=1.25 \invA, respectively.}
\label{fig4}
\end{figure*}

\begin{figure*}
\includegraphics [width=0.90\textwidth]{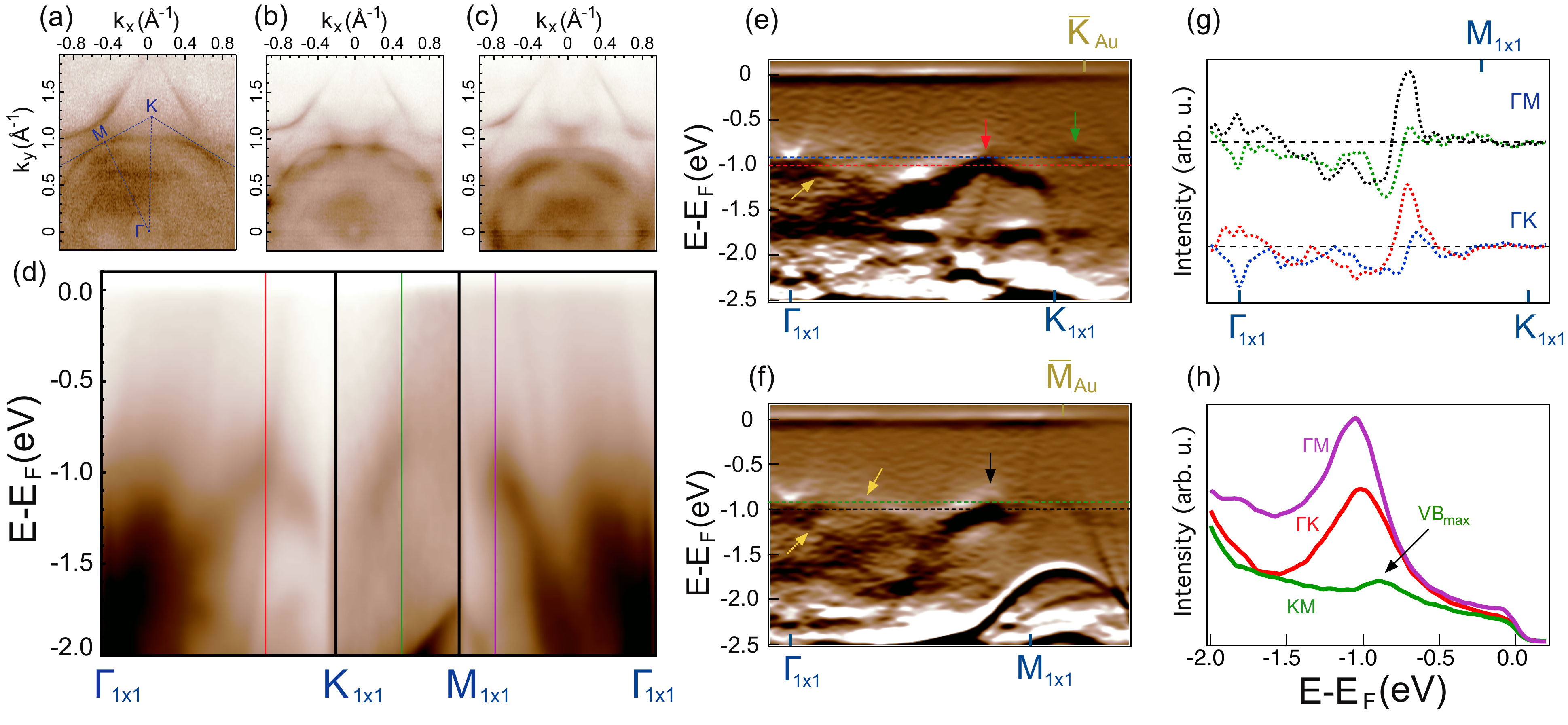}
\caption{Constant energy contours of (4x4)-BlueP on Au(111) at BEs (a) 0.025 eV (b) 0.85 eV and (c) 1.165 eV. Blue bashed lines in (a) border part of the first BZ of BlueP. (d) Band dispersions along the high-symmetry directions marked with blue dashed lines in (a). (e), (f) Second derivative plots of the ARPES spectra along the \GbarKbarBlueP, \GbarMbarBlueP\ directions, respectively. All ARPES spectra were acquired using horizontally polarized light with h$\nu = 60$ eV. (g) Momentum distribution curves extracted from (e), ((f)) at BEs of 0.92 eV using blue dashed line (green dashed line) and 1 eV red dashed line (black dashed line), respectively. (h) Energy distribution curves at the momentum positions marked in (d).}
\label{fig5}
\end{figure*}

Figures \ref{fig4} (b), (d)-(f) (Figs. \ref{fig5} (e), (f)) show the electronic signature of (4x4)-BlueP on Au(111), grown with the post-deposition annealing method (and with the Au(111) substrate at 230 $^{\circ}$C during P deposition), along the two inequivalent high-symmetry directions of the hexagonal BZ of BlueP, namely \GbarKbarBlueP\ and \GbarMbarBlueP. The band dispersions are presented as second derivate plots in order to disentangle the P-related bands from the intense Au(111)-derived features. Upon the formation of the (4x4)-BlueP,  dispersive bands are well discernible in the region between the Fermi level and 2 eV BE, where the Au sp-bands have relatively low intensity, in accordance with the DFT calculations (see Fig. \ref{fig2} (f)) while below 2 eV BE it is difficult to distinguish P-related electronic features due to the high density of states of the Au d-bands. We clearly distinguish hole-like bands dispersing towards the high-symmetry points of the BlueP BZ, namely K and M, see Figs. \ref{fig4} (b), (d) (and Fig. \ref{fig5} (e), (f)) where the bands are marked with red and black arrows, respectively. These pronounced P bands exhibit a two-dimensional character as they do not disperse with different excitation energies, see Fig. \ref{fig4} (g) (see also Fig. S4), which is consistent with the two-dimensional nature of (4x4)-BlueP. Additionally, in Figs. \ref{fig4} (b), (d)-(f) (Fig. \ref{fig5} (e), (f)) we observe bands with lower spectral weight dispersing towards the $\Gamma$ point (marked with yellow arrows), with their apex at the center of the surface BZ being higher in BE compared to the off-centered dispersive bands as one can conclude by the momentum distribution curves of Fig. \ref{fig4} (h) (and Fig. \ref{fig5} (g)). Notably, these faint electronic features seem consistent with their theoretically predicted band dispersions close to the BZ center as presented in Fig. \ref{fig2} (c).  Overall, the observed electronic bands share some resemblance with the theoretical calculations presented in Fig. \ref{fig2} (c) for freestanding BlueP indicating that the (4x4)-BlueP is linked with BlueP.

The main differences in the electronic structure of (4x4)-BlueP with the ideal freestanding BlueP are the following:
first, the apex of the off-center dispersive band along the \GbarKbarBlueP\ is found around 0.2 \invA\ away of the theoretically predicted value and towards the K point of BlueP, similarly, a comparable k-shift occurs for the band along the \GbarMbarBlueP\ direction towards the M high-symmetry point. Second, these bands (with a $p_z$ character as predicted by DFT) disperse towards the $\Gamma$ point of BZ to higher BE compared to the hole-like bands near the $\Gamma$ point (with a $p_{x, y}$ character) as seen in Figs. 5 (e), (f), in contrast to theoretical predictions where the $p_z$-like bands are lower in BE compared to the $p_{x, y}$ ones (see Fig. \ref{fig2} (c)). Finally, we observe an electronic feature that crosses the \Kbar\ point of the surface BZ of Au(111) (along the \KbarMbarBlueP\ high-symmetry direction of BlueP) as displayed in the close-up spectrum in Fig. \ref{fig4} (e) and highlighted by a green arrow. This state can be clearly seen along the equivalent high-symmetry path marked in Fig. \ref{fig5} (a) and displayed in Fig. \ref{fig5} (d). We readily observe a P-related electronic state that shifts the top of the VB of (4x4)-BlueP to the \KbarMbarBlueP\ high-symmetry direction of BlueP in stark contrast to theoretical predictions where the VB maximum is be located along the \GbarKbarBlueP\ of BlueP. The VB maximum of (4x4)-BlueP is found at 0.9 eV BE, which sets the lowest value for the band gap of the semiconducting BlueP (see Figs. \ref{fig4} (h), (i) and Figs. \ref{fig5} (g), (h)). If one overlooks the momentum offset of the VB's maximum, its energy position agrees with our DFT calculations as well as with scanning tunneling spectroscopy measurements \cite{Zhang:2016ex}. This serendipity can be attributed to the combined effect of the band gap modification of BlueP with strain\cite{Zhu:2014kf} and the n-doping of BlueP on Au(111)\cite{Zhang:2016ex}.

Since one does not expect that strong correlation effects play a role in the BlueP/Au(111) system, then, deviations from the DFT calculations should originate either from the (4x4) reconstruction of BlueP or from the interaction between BlueP and Au(111). Since we do not observe any sign of band back-folding from the (4x4) reconstruction on the electronic structure of BlueP (see a full photoemission mapping in Figs. \ref{fig5}(a)-(c)) we believe that the BlueP-Au(111) interaction is responsible for the observed changes in the electronic profile of BlueP. Therefore, we calculated the effective electronic structure of (4x4)-BlueP on Au(111) unfolded to the (1x1) BZ of the substrate using the BandUP code\cite{Madeiros:2014tg, Madeiros:2015eb}. It can be readily seen in Fig. \ref{fig4} (j), where colored arrows point to the corresponding dispersive features observed in ARPES, that the agreement between theory and experiment is restored if we consider the effective electronic structure of the BlueP-Au(111) system, indicating that the BlueP-Au(111) interaction drives the band renormalization of BlueP on Au(111). The band that sets the maximum of the VB lies along the \GbarKbar\ direction of Au(111) (and coincides with the \KbarMbarBlueP\ direction of BlueP) can be also seen in Fig. \ref{fig5} (d). The major differences between theory and ARPES occur for the bands with a $p_z$ character, indicating that the band renormalization is dictated by the interaction between the out-of-plane $p$ orbitals of (4x4)-BlueP and Au(111) surface. On the contrary, bands that disperse towards the center of the BZ, mainly with a $p_{x,y}$ character, do not renormalize considerably due to the weaker interaction with Au(111) and are in better agreement with the theoretical predictions for freestanding BlueP. Another effect that might enhance the interaction between Au(111) and BlueP might be the lone pair of electrons that any two-dimensional allotrope of P will have increasing the interactions with the proximate substrate. A potential mechanism that could drive the $p_z$ band renormalization might be the sublattice symmetry breaking imposed by the Au(111). The broken symmetry could also explain the VB maximum shift in (4x4)-BlueP along the \KbarMbarBlueP\ high-symmetry direction of BlueP. In Fig. \ref{fig2}(c) a Dirac-like dispersion can be readily seen near the K point of BlueP BZ at around 3 eV BE. The sublattice symmetry breaking can open a band gap shifting the upper part of the Dirac-like dispersion towards lower BEs\cite{Supplementary}. The upper part of the branch of this potentially gapped Dirac-like dispersion is observed in our ARPES spectra, while the lower one is buried in the high density of states of the Au d-bands and cannot be resolved. A similar effect has been observed in relevant two-dimensional structures such as (3x3)-Silicene on Ag(111), where the absence of the theoretically predicted Dirac fermions\cite{Cahangirov:2009fh} manifested as a gap opening at the Dirac point has been attributed to the substrate-induced sublattice symmetry breaking\cite{Lin2013rh}.

In summary, we report on the observation of the band renormalization of BlueP on Au(111). The combination of diffraction and microscopy measurements with theoretical calculations facilitated the identification of the structural details of the (4x4) BlueP reconstruction on Au(111). Most importantly, using APRES we were able to observe the momentum-resolved electronic signature of the graphene-like allotrope of P. Our study reveals an orbital-dependent band renormalization that is mainly linked to the interaction with the Au(111) substrate and not to the (4x4) lattice reconstruction, therefore, showing that the key for the realization of the ideal BlueP lattice is the reduction of P-substate interaction.

E. Golias would like to thank E. D. L. Rienks for helpful discussions.
The authors would like to thank the HPC Service of ZEDAT, Freie Universit\"at Berlin, for computing time.
We thank Helmholtz-Zentrum Berlin for the allocation of synchrotron radiation beamtime.

\bibliographystyle{aipnum4-1}

\end{document}